\documentclass[
aps, %
prl, floatfix, 
twocolumn, superscriptaddress, 10pt, 
a4paper, %
amsfonts, amsmath, amssymb]{revtex4-1} %
\usepackage[sort&compress]{natbib}
\usepackage{newtxtext}
\usepackage{txfonts}
\usepackage{ascmac}
\usepackage{mathrsfs}
\usepackage{bm}
\usepackage{color}
\usepackage{graphicx}
\usepackage{bm}
\usepackage{mathrsfs}
\usepackage[normalem]{ulem}
\usepackage{hhline}
\usepackage{hyperref}
\hypersetup{%
    unicode=false, %
    bookmarksnumbered=true, %
    bookmarkstype=toc, %
    breaklinks=true, %
    colorlinks=true, %
    pdfborder={0 0 1}, %
    linkcolor=blue, %
    citecolor=blue, %
    urlcolor=blue, %
    pdftoolbar=true, %
    pdfmenubar=true, %
    pdffitwindow=false, %
    pdfstartview={FitH} %
}
\newcommand{\br}{\bm{r}}
\newcommand{\bk}{\bm{k}}
\newcommand{\bp}{\bm{p}}
\newcommand{\bq}{\bm{q}}
\newcommand{\zi}{i}
\newcommand{\dd}[1]{\mathrm{d} #1\,}
\newcommand{\tr}{\mathrm{tr}\,}

\newcommand{\kB}{k_{\mathrm{B}} }
\newcommand{\me}{m_{\mathrm{e}}}

\begin{document}
%
\title{Alternating Current-induced Interfacial Spin-transfer Torque}
\date{\today}
\author{Junji Fujimoto}
\email[E-mail address: ]{junji@ucas.ac.cn}
\affiliation{Kavli Institute for Theoretical Sciences, University of Chinese Academy of Sciences, Beijing, 100190, China.}

\author{Mamoru Matsuo}%
\affiliation{Kavli Institute for Theoretical Sciences, University of Chinese Academy of Sciences, Beijing, 100190, China.}
\affiliation{RIKEN Center for Emergent Matter Science (CEMS), Wako, Saitama 351-0198, Japan.}
\affiliation{Advanced Science Research Center, Japan Atomic Energy Agency, Tokai, 319-1195, Japan.}
\begin{abstract}
We investigate an interfacial spin-transfer torque and $\beta$-term torque with alternating current~(AC) parallel to a magnetic interface.
We find that both torques are resonantly enhanced as the AC frequency approaches to the exchange splitting energy.
We show that this resonance allows us to estimate directly the interfacial exchange interaction strength from the domain wall motion.
We also find that the $\beta$-term includes an unconventional contribution which is proportional to the time derivative of the current and exists even in absence of any spin relaxation processes.
\end{abstract}
\maketitle

\paragraph{Introduction.---} 
A variety of physical phenomena arises near interfaces, such as spin-dependent transports~\cite{baibich1988,valet1993,miyazaki1995,moodera1995,gould2004,miwa2017}, interfacial magnetic phenomena~\cite{carcia1985,camley1989,engel1991,fert1980,levy1981,crepieux1998a}, and chiral/topological phenomena~\cite{bode2007,ferriani2008,meckler2009,heinze2011}, which have attracted attention from many years ago~\cite{hellman2017}.
Among these, the spin-dependent transport has been closely related to the aspect of not only fundamental physics but device application; especially the tunneling magnetoresistance~\cite{valet1993,miyazaki1995,moodera1995} impacted upon the invention of the magnetoresistive random access memory~\cite{zhu2006}.

The spin-dependent transports near the interfaces are important from the viewpoint of the understanding of recent developments in spintronics, such as the spin pumping effect~(SPE)~\cite{mizukami2001,tserkovnyak2002,simanek2003,tserkovnyak2004,takahashi2010,ohnuma2014,tatara2017} and the spin Seebeck effect~(SSE)~\cite{uchida2008,adachi2011}, because the mutual dependence between the magnetization dynamics and the spin-dependent transports is the key mechanism in various spin-dependent phenomena.
The two effects are the ways of generating spin currents without electric currents, in a bilayer system consisting of a ferromagnet~(FM) and a normal metal~(NM); the spin precession due to the rf microwave in FM induces the spin current in NM in the case of SPE, and the temperature difference between FM and NM induces that for SSE.
Both of effects can be described by the tunnel Hamiltonian method~\cite{adachi2011,ohnuma2014,matsuo2018}, which also captures tunneling magnetoresistance.

The interfacial exchange interaction between conduction electrons in NM and magnetization in FM plays a crucial role in SPE and SSE, which are proportional to $J_{sd}^2$, where $J_{sd}$ is the interfacial exchange interaction strength~\cite{adachi2011,ohnuma2014}.
In general, the exchange interaction possibly gives rise to an essential contribution to spin-related phenomena near the interfaces, such as the spin Hall magnetoresistance~\cite{nakayama2013}.
However, this physically essential parameter $J_{sd}$ has not been directly measured, and the direct method of evaluating it is not yet proposed.

\begin{figure}[t]
\centering
\includegraphics[width=0.9\linewidth]{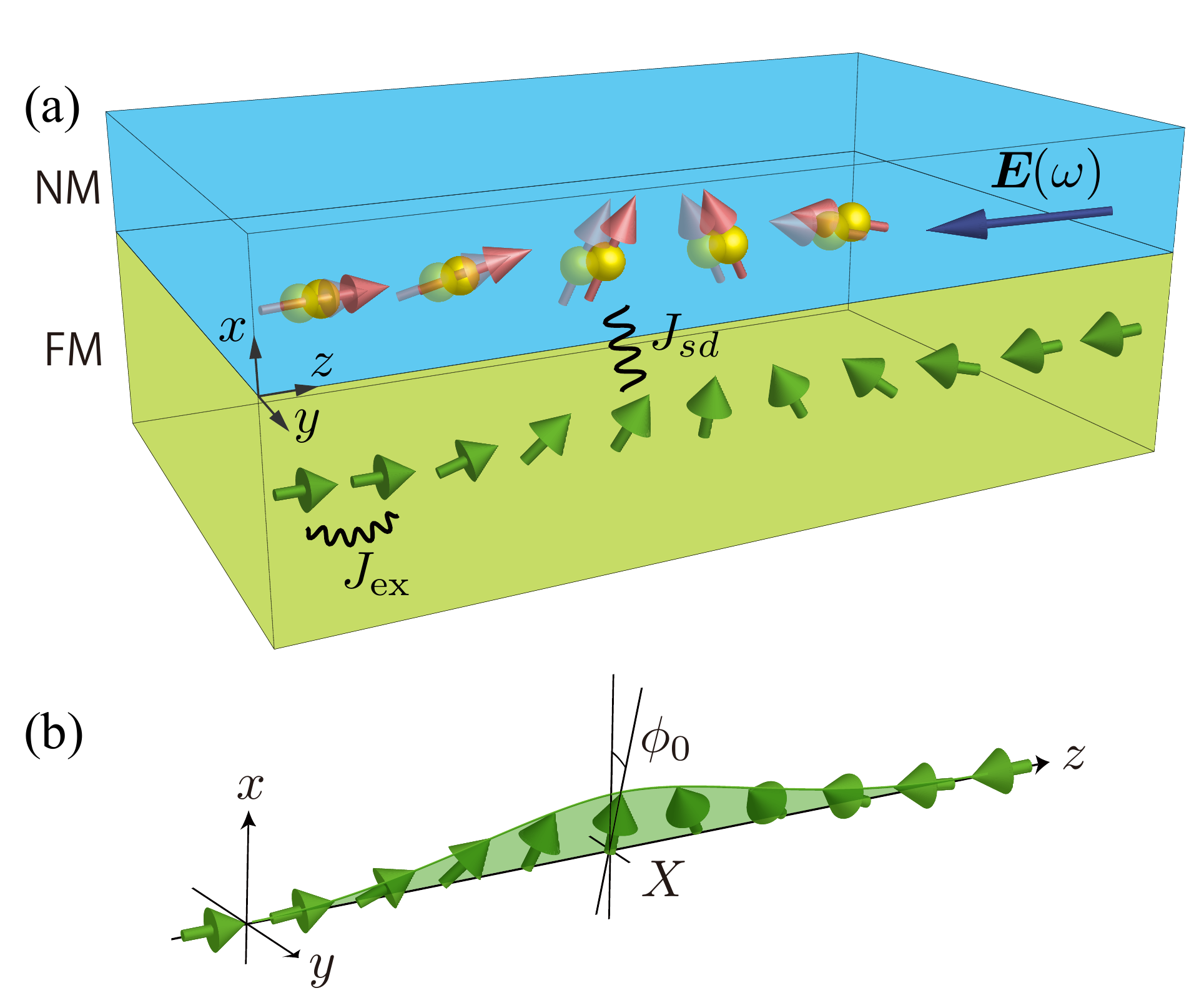}
\caption{\label{fig:1}%
(a)~Schematic description of the configuration, where a ferromagnet~(FM) having two magnetic domains with one domain wall~(DW) is adjoined by a normal metal~(NM) whose conduction electron couples with the magnetization in FM through the interfacial exchange interaction.
The alternating current~(AC) is applied parallel to the interface due to the electric field with finite frequency.
The AC-induced spin torques on the DW leads to the oscillation of the position of the DW center, which allows us to evaluate the interfacial exchange interaction strength.
(b)~DW configuration, which is described by the corrective coordinates of the DW center $X$ and the angle $\phi_0$.
}
\end{figure}

In this Letter, we present a direct method of evaluating the interfacial exchange interaction strength $J_{sd}$ from the domain wall dynamics in FM adjoined by NM, applying an alternating current~(AC) parallel to the interface~(Fig.~\ref{fig:1} (a)).
 It is a well-known fact that in bulk ferromagnetic metals with noncolinear magnetic textures such as domain walls, the direct current~(DC) accompanying with spin polarization exerts spin torques on the magnetization, which leads to its dynamics such as the domain wall motion~\cite{berger1984,berger1992,salhi1993,yamaguchi2004,zhang2004,tatara2008a}.
We here extend the DC-induced spin torques into the region of an arbitrary frequency of the current, based on the quantum field theoretical approach, and apply this to the interfacial exchange interacting system of the FM-NM bilayer.
We consider so thin NM that we focus only on the spin polarized electronic states near the interface due to the interfacial exchange interaction.
We find that the AC-induced spin torques consist of corresponding extensions of the spin-transfer torque~\cite{slonczewski1996,bazaliy1998,tatara2004,zhang2004} and the so-called $\beta$-term torque~\cite{thiaville2004,zhang2004,tserkovnyak2006,duine2007,kohno2007}.
However, we also find that the results we obtain include physically a novel contribution to the $\beta$-term torque, which depends on the time derivative of the current density.
Our important finding is that both spin torques are proportional to $(1 - \omega^2 \tau_{\mathrm{sd}}^2)^{-1}$ for the case of no spin relaxation processes, where $\omega$ is the AC frequency and $\tau_{\mathrm{sd}} = \hbar / 2 \Delta$ with $2 \Delta$ being the interfacial exchange splitting.
The exchange splitting is related to $J_{sd}$ by $\Delta = S J_{sd}$, where $S$ is the localized spin length constructing the magnetization.
This dependence suggests that we can evaluate $J_{sd}$ from the magnetization dynamics driven by the spin torques.
In the viewpoint of application, the enhancement of the spin torques have an advantage in that less current density is needed to excite the magnetization dynamics.

We then solve the equation of motion of a rigid domain wall~(DW)~\cite{tatara2004,tatara2008a} driven by the obtained spin torques, in the presence of a spin relaxation process.
The equation is expressed by the two collective coordinates; the position of the DW center $X$ and the angle $\phi_0$~(Fig.~\ref{fig:1} (b)), and the spin torques act as the forces to $X$ and $\phi_0$.
We find that $X$ and $\phi_0$ oscillate along with the frequency $\omega$ in the region of the small electric current density, and the amplitude of the oscillation of $X$ increases resonantly near $\omega \tau_{sd} \simeq 1$.
Hence, we conclude that the dependence of $X$ on the frequency allows us to estimate the interfacial exchange splitting.

This Letter is organized as follows.
We first present the total Lagrangian of the magnetization in FM and the conduction electron in NM as well as their interfacial exchange interaction, and introduce the \textit{rotated frame picture} sometimes used in the context of the ferromagnetic spintronics.
Then, the AC-induced spin torques are evaluated based on the linear response theory with the thermal Green function method.
As an application, we consider the DW dynamics driven by the obtained spin torques.
\paragraph{Theory.---} 
The total Lagrangian that we consider is given by $\mathcal{L} = \mathcal{L}_{m} + \mathcal{L}_{e} - \mathcal{H}_{sd}$, where $\mathcal{L}_{m}$ is the Lagrangian of the magnetization in the FM layer, $\mathcal{L}_{e}$ is that of the conduction electron in the NM layer, and $\mathcal{H}_{sd}$ is the $sd$-like interfacial exchange interaction between them.

Considering that the magnetization is constructed by the localized spins ordering, we express the Lagrangian of the magnetization as that of the localized spin, $\bm{M} = - M_S \bm{m}$ with $\bm{m} = (\sin \theta \cos \phi, \sin \theta \sin \phi, \cos \theta)$, where $M_S$ is the saturated magnetization, and $\theta = \theta (\br, t)$, $\phi = \phi (\br, t)$.
Here, $\bm{m}$ does not represent the unit vector of the magnetization, but that of the localized spin, whose signs are opposite.
The Lagrangian of the localized spin is defined as $\mathcal{L}_{m} = \int \mathrm{d} \br \,(\hbar S / a^3) \dot{\phi} (\cos \theta - 1) - \mathcal{H}_m$ with
\begin{align}
\mathcal{H}_m
	& = \int \frac{\mathrm{d} \br}{a^3} \left[ \frac{ J_{\mathrm{ex}} }{2} S^2 \left( \bm{\nabla} \bm{m} \right)^2 - \frac{K}{2} S^2 m_z^2 + \frac{K_{\perp}}{2} S^2 m_y^2 \right]
,\end{align}
where $a$ is the lattice constant of FM, $J_{\mathrm{ex}}$ is the exchange interaction between the localized spins, and $K$ and $K_{\perp}$ are easy- and hard-axis magnetic anisotropies, respectively.
Note that the saturated magnetization $M_S$ is related to the localized spin length by $M_S = \gamma_e \hbar S / a^3$ with the gyromagnetic ratio $\gamma_{\mathrm{e}}$, and $J_{\mathrm{ex}}$, $K$, and $K_{\perp}$ are all positive.

We show the rest of the Lagrangian, which is written by $\mathcal{L}_{e} - \mathcal{H}_{sd} = \int \mathrm{d} \bm{r} \psi^{\dagger} (\br, t) ( \zi \hbar \partial_t - H_e - H_{sd} ) \psi^{} (\br, t)$, where $\psi^{(\dagger)}$ is the field operator of electrons, $H_e = \bp^2/2 \me + V$ describes the kinetic energy with the electron mass $\me$ and the nonmagnetic and magnetic impurity potentials given by $V = u_{\mathrm{i}} \sum_{i=1}^{N_{\mathrm{i}} } \delta (\br - \bm{R}_i) + u_{\mathrm{s}} \sum_{j=1}^{N_{\mathrm{s}} } ( \bm{S}_{\mathrm{imp}, j} \cdot \bm{\sigma} ) \delta (\br - \bm{R}'_j)$ with the impurity numbers $N_{\mathrm{i}}$ and $N_{\mathrm{s}}$ and with the strengths $u_{\mathrm{i}}$ and $u_{\mathrm{s}}$, and $H_{sd} = - \Delta \bm{m} (\br,t) \cdot \bm{\sigma}$ represents the interfacial exchange interaction with the coupling constant $\Delta > 0 $ with $\bm{\sigma} = (\sigma^x, \sigma^y, \sigma^z)$ being the Pauli matrices.
The magnetic impurity spin $\bm{S}_{\mathrm{imp},j}$ is assumed to be quenched.

Then, we transform the Hamiltonian into the `rotated frame'~\cite{korenman1977,tatara2008a} by using the unitary transformation $U (\br, t)$ defined by $U^{\dagger} (\br, t) \left( \bm{m} (\br, t) \cdot \bm{\sigma} \right) U (\br, t) = \sigma^z$ with $\bar{\psi} = U^{\dagger} \psi$.
The physical meaning of the unitary transformation is that the quantization axis of the electron spin is to be reoriented to $\bm{m} (\br, t)$ at each position and time.
Hence, we call the frame after the transformation as the rotated frame and denote $\bar{A}$ as the quantity $A$ in the rotated frame.
The electron described by $\bar{\psi}^{(\dagger)}$ feels the uniform exchange interaction in the rotated frame.
We also express the rotational unitary transformation by using the rotational matrix $\mathcal{R} (\br, t)$ for the three-dimensional vector defined by $U^{\dagger} (\br, t) \, \bm{\sigma} \, U (\br, t) = \mathcal{R} (\br, t) \bm{\sigma}$.
This expression of the unitary transformation is useful for the magnetic impurity potential and the spin torques.
Note that the relation to the definition of $U$ is $U^{\dagger} (\bm{m} \cdot \bm{\sigma} ) U = \bm{m} \cdot (\mathcal{R} \bm{\sigma}) = (\mathcal{R}^{-1} \bm{m}) \cdot \bm{\sigma} = \sigma^z$, hence $\mathcal{R}^{-1} \bm{m} = \hat{z}$, where $\hat{z}$ is the unit vector along the $z$-axis.

We now look into the equation of motion of the localized spin, which is obtained from the Euler-Lagrange equation with the relaxation function $\mathcal{W}$~\cite{tatara2008a},
\begin{align}
\frac{ \mathrm{d} }{ \mathrm{d} t } \left( \frac{\delta \mathcal{L}}{\delta \dot{q}} \right) - \frac{\delta \mathcal{L}}{\delta q}
	& = - \frac{\delta \mathcal{W}}{\delta \dot{q}}
\label{eq:EL}
,\end{align}
where $q \in \{\theta, \phi\}$, and $\mathcal{W} = \int \mathrm{d} \br\, (\hbar S \alpha_{\mathrm{G}}/2 a^3) \dot{\bm{m}}^2$ with the Gilbert damping constant $\alpha_{\mathrm{G}}$.
Equation~(\ref{eq:EL}) leads to the Landau-Lifshitz-Gilbert equation,
$\dot{\bm{m}}
	= \gamma_{\mathrm{e}} \bm{H}_{\mathrm{eff}} \times \bm{m}
	+ \alpha_{\mathrm{G}} \dot{ \bm{m} } \times \bm{m}
	+ \bm{\tau}_{e}
,$ where $\bm{H}_{\mathrm{eff}}$ is the effective magnetic field defined as $\gamma_{\mathrm{e}} \bm{H}_{\mathrm{eff}} = (1/\hbar S)\, \delta \mathcal{H}_m / \delta \bm{m}$, and $\bm{\tau}_e$ is the spin torque through the interfacial exchange interaction;
\begin{align}
&\gamma_{\mathrm{e}} \bm{H}_{sd}
	= \frac{1}{\hbar S} \left\langle \frac{\delta \mathcal{H}_{sd}}{\delta \bm{m}} \right\rangle_{\mathrm{neq}}
, \\
&\bm{\tau}_{e}
	= \gamma_{\mathrm{e}} \bm{H}_{sd} \times \bm{m}
	= - \frac{\Delta}{\hbar S} \left\langle\bm{s}\right\rangle_{\mathrm{neq}} \times \bm{m}
\label{eq:tau_e}
.\end{align}
Here, $\bm{s} = \bm{s} (\br, t) = \psi^{\dagger} (\br, t) \bm{\sigma} \psi (\br, t)$ is the spin density operator divided by $\hbar / 2$, and $\langle\,\cdots\rangle_{\mathrm{neq}}$ describes the statistical average in the nonequilibrium.

The spin torque is expressed in the rotated frame as $\bar{\bm{\tau}}_{e} = \mathcal{R}^{-1} \bm{\tau}_{e} = - (\Delta/\hbar S) \langle \bar{\bm{s}} \rangle_{\mathrm{neq}} \times \hat{z}$.
We emphasise that, in the rotated frame, the perpendicular components of the nonequilibrium spin polarization $\langle \bar{\bm{s}} \rangle_{\mathrm{neq}}$ to the $\hat{z}$-axis only act as torques.

In this Letter, we evaluate the nonequilibrium spin polarization $\langle \bar{\bm{s}} \rangle_{\mathrm{neq}}$ in the linear response to the electric field $\bm{E} (\omega)$ with the frequency $\omega$ as
\begin{align}
\langle \bar{s}^{\alpha} (\bq, \omega) \rangle_{\mathrm{neq}}
	& = \bar{\chi}^{\alpha i} (\bq, \omega) E_i (\omega)
\label{eq:spin_in_rotated_frame}
\end{align}
in the Fourier space ($\alpha = x, y$ and $i = y, z$).
From the linear response theory, we can obtain the response coefficient $\bar{\chi}^{\alpha i} (\bq, \omega)$ from
$\bar{\chi}^{\alpha i} (\bq; \omega) = (\bar{K}^{\alpha i} (\bq; \omega) - \bar{K}^{\alpha i} (\bq; 0)) / \zi \omega$, where $\bar{K}^{\alpha i} (\bq; \omega)$ can be evaluated from the following spin-current correlation function in the Matsubara form
\begin{align}
\bar{\mathcal{K}}^{\alpha i} (\bq; \zi \omega_{\lambda})
	& = \int_0^{1/\kB T} \dd{\tau} e^{\zi \omega_{\lambda} \tau} \left\langle \mathrm{T}_{\tau} \bar{s}^{\alpha} (\bq, \tau) \bar{J}_i (0) \right\rangle_{\mathrm{eq}}
\label{def:correlation_in_Matsubara}
\end{align}
through the analytic continuation $\zi \omega_{\lambda} \to \hbar \omega + \zi 0$; $\bar{K}^{\alpha i} (\bq; \omega) = \bar{\mathcal{K}}^{\alpha i} (\bq; \hbar \omega + \zi 0)$.
Here $\omega_{\lambda} = 2 \pi \lambda \kB T$ is the Matsubara frequency of bosons with the temperature $T$~\cite{fujimoto2019}, and the spin and the electric current operator in the rotated frame are given by 
\begin{align}
\bar{s}^{\alpha} (\bq)
	& = \frac{1}{V} \sum_{\bk} \bar{c}^{\dagger}_{\bk-\bq} \sigma^{\alpha} \bar{c}^{}_{\bk}
\label{eq:spin}
,\end{align}
\begin{align}
\bar{J}_i
	& = - e \sum_{\bk} \frac{\hbar k_i}{\me} \bar{c}^{\dagger}_{\bk} \bar{c}^{}_{\bk}
		- \frac{e \hbar}{2 \me} \sum_{\bq'} A^{\beta}_{i} (\bq') \sum_{\bk} \bar{c}^{\dagger}_{\bk} \sigma^{\beta} \bar{c}^{}_{\bk-\bq'}
\label{eq:current}
,\end{align}
where $\bar{c}^{(\dagger)}_{\bk}$ is the Fourier transform of the field operator $\bar{\psi}^{(\dagger)} (\br)$.
The first term of Eq.~(\ref{eq:current}) is the normal velocity term and the second is the anomalous velocity term due to the spin gauge field $A_i^{\beta} (\br) = - \zi \tr [ U^{\dagger} \partial_i U \sigma^{\beta} ]$ with $\beta = x, y, z$.

As the detailed calculation will be shown elsewhere, we here sketch out the procedures of the calculation.
By substituting Eqs.~(\ref{eq:spin}) and (\ref{eq:current}) into Eq.~(\ref{def:correlation_in_Matsubara}), we rewrite the correlation function by using the thermal Green functions according to Wick's theorem.
We expand the Green function by the spin gauge field up to the first order, and take the statistical average on the impurity positions, then we obtain
\begin{align}
\bar{\mathcal{K}}^{\alpha i} (\bq; \zi \omega_{\lambda})
	& = \frac{e \hbar}{4 \me V} A^{\beta}_{j}(\bq) \kB T \sum_{n} \sum_{\bk} \Phi_{i j}^{\alpha \beta} (\bk; \zi \epsilon_n^{+}, \zi \epsilon_n)
\end{align}
where $\epsilon_n^{+} = \epsilon_n + \omega_{\lambda}$, $\epsilon_n = (2 n + 1) \kB T$ is the Matsubara frequency of fermions, and
\begin{align}
\Phi_{i j}^{\alpha \beta} (\bk; \zi \epsilon_n^{+}, \zi \epsilon_n)
	& = 2 \delta_{i j} \tr \left[
			\Lambda^{\alpha} g_{\bk} (\zi \epsilon_n^{+}) \sigma^{\beta} g_{\bk} (\zi \epsilon_n)
		\right]
\nonumber \\ & \hspace{0em}
		- \frac{\hbar^2 k_i k_j}{\me}
		\tr \left[
			\Lambda^{\alpha} g_{\bk} (\zi \epsilon_n^{+}) \sigma^{\beta} g_{\bk} (\zi \epsilon_n^{+}) g_{\bk} (\zi \epsilon_n)
		\right]
\nonumber\\ & \hspace{0em}
		- \frac{\hbar^2 k_i k_j}{\me}
		\tr \left[
			\Lambda^{\alpha} g_{\bk} (\zi \epsilon_n^{+}) g_{\bk} (\zi \epsilon_n) \sigma^{\beta}  g_{\bk} (\zi \epsilon_n)
		\right]
\label{def:Phi}
.\end{align}
Here, $g_{\bk} (\zi \epsilon_n^{(+)}) = (\zi \epsilon_n^{(+)} + \mu - \hbar^2 k^2 / 2 \me - \Delta \sigma^z - \Sigma (\zi \epsilon_n^{(+)}) )^{-1}$ is the thermal Green function with the self energy within the self-consistent Born approximation $\Sigma ( \zi \epsilon_n) = n_{\mathrm{i}} u_{\mathrm{i}}^2 \sum_{\bk} g_{\bk} (\zi \epsilon_n) + n_{\mathrm{s}} u_s^2 S_{\mathrm{imp}}^2 \sum_{\bk} \sigma^{\gamma} g_{\bk} (\zi \epsilon_n) \sigma^{\gamma}$, where $n_{\mathrm{i}}$ and $n_{\mathrm{s}}$ are the impurity concentrations of nonmagnetic and magnetic impurities, respectively, and we have taken the statistical average on the impurity spins and assume the spherical spins, $\overline{S^{\alpha}_{\mathrm{imp}, i} S^{\beta}_{\mathrm{imp},j}} = (S_{\mathrm{imp}}^2/3) \delta_{i j} \delta^{\alpha \beta}$.
In Eq.~(\ref{def:Phi}), we have evaluated $\Phi_{i j}^{\alpha \beta}$ by assuming $\bq = 0$ since $\bar{\mathcal{K}}^{\alpha i} (\bq; \zi \omega_{\lambda})$ is already in the $q$-linear order because of $A_i^{\beta} (\bq)$.
The full vertex of spin $\Lambda^{\sigma} = \Lambda^{\sigma} (\zi \epsilon_n^{+}, \zi \epsilon_n)$ is given by
$ \Lambda^{\alpha}
	 = \sigma^{\alpha}
		+ n_{\mathrm{i}} u_{\mathrm{i}}^2 \sum_{\bk} g_{\bk} (\zi \epsilon_n^{+}) \Lambda^{\alpha} g_{\bk} (\zi \epsilon_n)
		+ \frac{1}{3} n_{\mathrm{s}} u_{\mathrm{s}}^2 S_{\mathrm{imp}}^2\sum_{\bk} \sigma^{\gamma} g_{\bk} (\zi \epsilon_n^{+}) \Lambda^{\alpha} g_{\bk} (\zi \epsilon_n)  \sigma^{\gamma}$.
After some straightforward calculation and taking the analytic continuation with the assumption of $T = 0$, we then obtain
\begin{align}
\bar{\chi}^{\alpha i} (\bq, \omega)
	& = \frac{\hbar \sigma_{\mathrm{s}} }{e}A^{\beta}_i (\bq) \sum_{\sigma = \pm} \left( \delta^{\alpha \beta} + \zi \sigma \epsilon^{\alpha \beta z} \right)
	\frac{ \sigma }{ 2 \sigma \Delta - \hbar \omega + \zi \hbar / \tau_{\mathrm{s}} }
\label{eq:result:chi}
,\end{align}
where we neglected the higher order contribution of $\hbar / \epsilon_{\mathrm{F} \sigma} \tau_{\sigma}$ with the spin-dependent Fermi energy $\epsilon_{\mathrm{F} \sigma}$  and the momentum lifetime $\tau_{\sigma}$.
Here, $\sigma_{\mathrm{s}} = \sigma_{\mathrm{s}} (\omega)$ is the spin conductivity, $ \sigma_{\mathrm{s}} (\omega) = (e^2/\me) \sum_{\sigma = \pm} \sigma n_{\sigma} \tau_{\sigma} / (1 - \zi \omega \tau_{\sigma}) $, with the spin-dependent electron density $n_{\sigma}$ and lifetime $\tau_{\sigma}$ with $\sigma = \pm$, and $\tau_s$ is the relaxation time due to the magnetic impurity scattering defined as $\tau_{\mathrm{s}}^{-1} = (2 \pi / 3 \hbar) n_{\mathrm{s}} u_{\mathrm{s}}^2 S_{\mathrm{imp}}^2 \bigl( \nu_{+} +\nu_{-} \bigr)$, where $\nu_{\pm}$ is the density of states at the Fermi level.
\paragraph{Results.---} 
Here we show the expression of the AC-induced spin-transfer torque and $\beta$-term torque obtained from Eq.~(\ref{eq:result:chi}) combined with Eq.~(\ref{eq:spin_in_rotated_frame}) and (\ref{eq:tau_e}),
\begin{align}
\bm{\tau}_{e} (\br, \omega)
	& =
	\frac{
		\bigl( \bm{j}_{\mathrm{s}} \cdot \bm{\nabla} \bigr) \bm{m}
		+ (\zi \omega \tau_{sd} + \zeta_{\mathrm{s}}) \bm{m} \times \bigl( \bm{j}_{\mathrm{s}} \cdot \bm{\nabla} \bigr) \bm{m}
	}{ 1 + (\zi \omega \tau_{sd} + \zeta_{\mathrm{s}})^2 }
\label{eq:result:torque}
\end{align}
in the laboratory frame, where $\tau_{sd} = \hbar / 2 \Delta$, $\zeta_{\mathrm{s}} = \tau_{sd} / \tau_{\mathrm{s}}$, and we used $\mathcal{R} \bm{A}^{\perp}_i = - \bm{m} \times \partial_i \bm{m}$ and $\mathcal{R} (\hat{z} \times \bm{A}^{\perp}_i) = \partial_i \bm{m}$.
The frequency-dependent spin current is denoted by $\bm{j}_{\mathrm{s}} = \bm{j}_{\mathrm{s}} (\omega) = (\hbar/2eS) \sigma_{\mathrm{s}} (\omega) \bm{E} (\omega)$.
By taking the static field limit of $\omega \to 0$, we find that the first term proportional to $\bigl( \bm{j}_{\mathrm{s}} \cdot \bm{\nabla} \bigr) \bm{m}$ corresponds to the spin-transfer torque and the second term proportional to $\bm{m} \times \bigl( \bm{j}_{\mathrm{s}} \cdot \bm{\nabla} \bigr) \bm{m}$ coincides with the $\beta$-term torque, and we confirm that our result agrees with that by Zhang and Li~\cite{zhang2004} and by Kohno and Shibata~\cite{kohno2007} for the model of the conduction electron in a ferromagnet, although they are not for the interfacial exchange interaction like our situation.
Hence, our result~(\ref{eq:result:torque}) is an extension of the DC-induced spin torques into the arbitrary frequency.
Equation~(\ref{eq:result:torque}) is the main result of this Letter.

 Considering the case of the dilute magnetic-impurity concentration, $\tau_{\mathrm{s}} \to \infty$, so that $\zeta_{\mathrm{s}} \to 0$, we find
\begin{align*}
\bm{\tau} (\br, \omega)
	& = \frac{ 1 }{ 1 - \omega^2 \tau_{sd}^2 }
		\left( ( \bm{j}_{\mathrm{s}} \cdot \bm{\nabla} ) \bm{m}
		+ \zi \omega \tau_{sd} \bm{m} \times \bigl( \bm{j}_{\mathrm{s}} \cdot \bm{\nabla} \bigr) \bm{m} \right)
,\end{align*}
which implies that the spin torques increase resonantly as the AC frequency approaches to the $1/\tau_{sd}$.
As shown in \textit{Application}, this frequency dependence allows us to determine the magnitude of the interfacial exchange interaction.

We also find that the $\beta$-term toque is present proportional to the frequency $\omega$, without the magnetic impurity which results in a spin relaxation process.
The $\beta$-term torques are known to arise from the spin relaxation process~\cite{zhang2004}, such as the scattering due to the magnetic impurity potential~\cite{kohno2006} and spin-orbit impurity potential~\cite{tatara2008}.
Actually, Eq.~(\ref{eq:result:torque}) shows that there is also the $\beta$-term proportional to the magnetic impurity concentration, $\zeta_{\mathrm{s}} \sim n_{\mathrm{s}}$.
The $\beta$-term torques also arise from \textit{nonadiabaticity}, which stands for the higher order of the derivatives, such as the terms proportional to $\bm{m} \times \partial_t \partial_i \bm{m}$~\cite{tserkovnyak2006,thorwart2007}.
Note that Eq.~(\ref{eq:result:torque}) is the Fourier form in the frequency space, so that in the real time space, the $\beta$-term torque is expressed as $\bm{m} \times \bigl( (\mathrm{d} \bm{j}_{\mathrm{s}} / \mathrm{d} t) \cdot \bm{\nabla} \bigr) \bm{m}$, which is the first order derivative for the magnetization, not higher orders.
For these reasons, the $\beta$-term torque we obtain is different from the ones which are already known.

It should be discussed that the relation of the spin torques obtained here to the Rashba spin-orbit torques~(SOT)~\cite{manchon2008,obata2008,manchon2009,kurebayashi2014} and the spin Hall torques~(SHT)~\cite{kim2013b,garello2013,pai2015}.
Both Rashba SOT and SHT originate from the spin-orbit couplings~(SOCs); Rashba SOT comes from the interfacial SOC due to the inversion symmetry breaking and SHT arises from the bulk SOC in NM.
We have assumed that these SOCs are weak so that these torques do not contribute much; for instance, that is the case of Cu as a NM and Py as a FM.
For the strong SOCs, we have to develop our theory which contains these strong SOCs, but that is out of our focus.

\begin{figure}[tbp]
\centering
\includegraphics[width=0.9\linewidth]{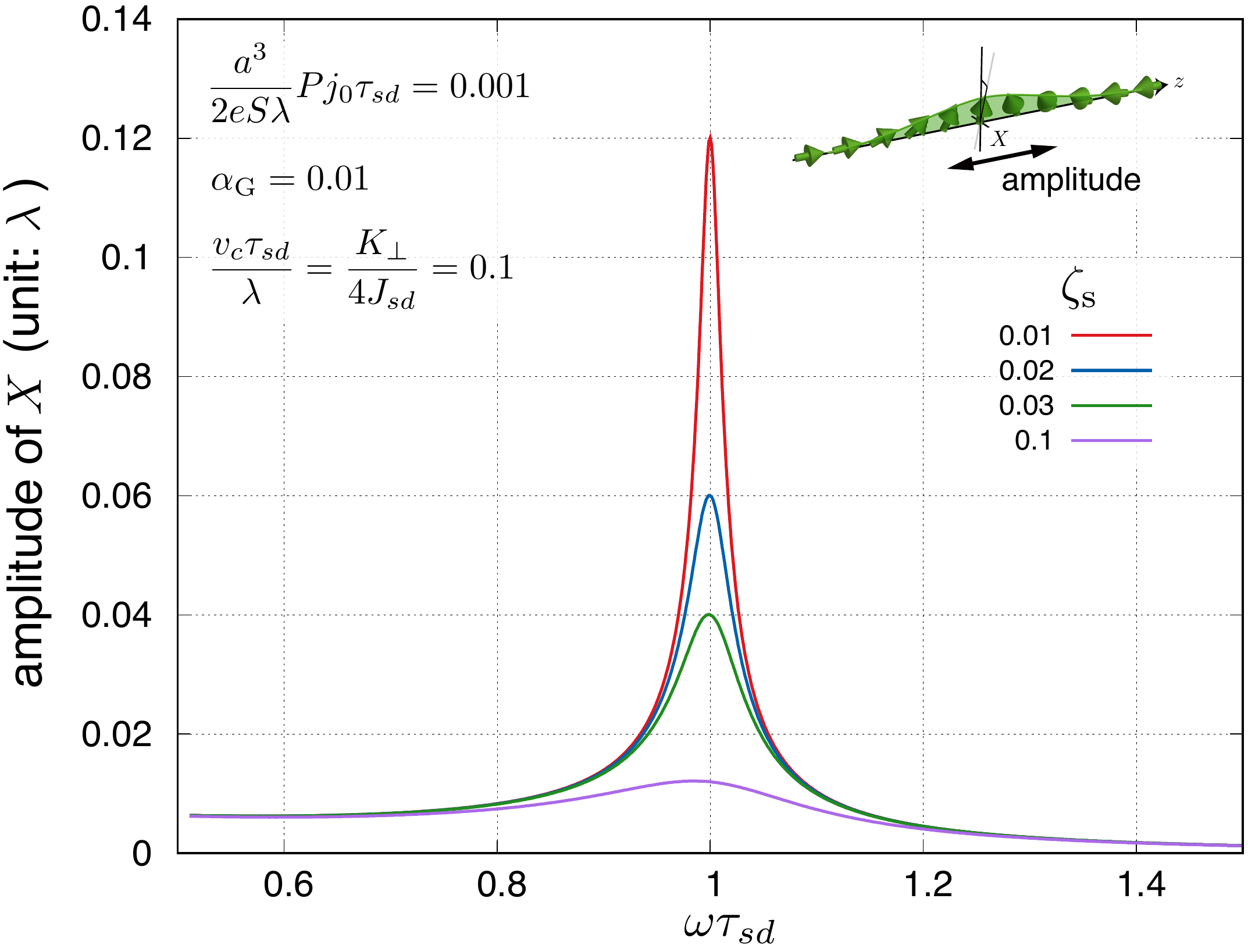}
\caption{\label{fig:2}%
Amplitude of the oscillation of the position $X$ during the period $1 / 2 \pi \omega$ for various $\zeta_{\mathrm{s}}$.
It is clear that the amplitude enhances near the resonance point $\omega \tau_{sd} = 1$.
}
\end{figure}

\paragraph{Application.---} 
 Now, we focus on the domain wall~(DW) motion as an application of the obtained torques.
 Following Tatara \textit{et al.}~\cite{tatara2008a} and assuming $K_{\perp} \ll K$ and no pinning potentials, we rewrite the Lagrangian $\mathcal{L}_m$ into that of DW, introducing the corrective coordinates of the DW center $X(t)$ and the angle $\phi_0(t)$~(Fig.~\ref{fig:1} (b)),
\begin{align}
\bm{m}
	& = \left(
		\frac{\cos \phi_0 (t)}{\cosh \frac{z - X(t)}{\lambda}}, \quad
		\frac{\sin \phi_0 (t)}{\cosh \frac{z - X(t)}{\lambda}}, \quad
		\tanh \frac{z - X(t)}{\lambda}
	\right)
,\end{align}
where $\lambda = \sqrt{J_{\mathrm{ex}} / K}$ is the DW width.
By using $X(t)$ and $\phi_0(t)$, the DW Lagrangian and the dissipation function are written by $\mathcal{L}_{\mathrm{w}} = N_{\mathrm{w}} S \left( \hbar \dot{X} \phi_0 / \lambda - (K_{\perp} S / 2) \sin^2 \phi_0 \right)$ and $\mathcal{W}_{\mathrm{w}} = \frac{\alpha N_{\mathrm{w}} \hbar S}{2} \left( \frac{\dot{X}^2}{\lambda^2} + \dot{\phi}_0^2 \right)$, where $N_{\mathrm{w}} = 2 \lambda A / a^3$ is the number of spins in the wall with $A$ being the cross-sectional area.
We have neglected the spin wave excitations.
From these, the equation of motion is written as
\begin{subequations}
\begin{align}
\dot{\phi}_0 + \alpha \frac{\dot{X}}{\lambda}
	& = - \tau_{sd} \frac{ \mathrm{d} \mathcal{T} }{ \mathrm{d} t} + \zeta_{\mathrm{s}} \mathcal{T}
\label{eq:EOMa}
, \\
\frac{\dot{X}}{\lambda} - \alpha \dot{\phi}_0
	& = \frac{v_{c}}{\lambda} \sin 2 \phi_0
		+ \mathcal{T}
\label{eq:EOMb}
,\end{align}
where 
\begin{align}
\mathcal{T} = \frac{a^3}{2 e S \lambda} \frac{ P j_0 e^{-\zi \omega t} }{1 + (\zi \omega \tau_{sd} + \zeta_{\mathrm{s}})^2}
, \qquad
v_c = \frac{K_{\perp} \lambda S}{2 \hbar}
\end{align}
\end{subequations}
with the electric current density $j_0$ and its polarization $P$.
Here, $\mathcal{T}$ is the spin torques that we obtain and act as the forces to $X (t)$ and $\phi_0 (t)$.
Solving Eqs.~(\ref{eq:EOMa}) and (\ref{eq:EOMb}) numerically, we find that the DW position $X(t)$ and angle $\phi_0(t)$ oscillate with the period $2 \pi / \omega$ for the low current density $(a^3/2 e S) P j_0 \lesssim v_c$.
We also find that the amplitude of the oscillations become larger as $\omega \tau_{sd}$ approaching to unitary~(Fig.~\ref{fig:2}).
Figure~\ref{fig:2} depicts the oscillation amplitude of the DW position for the case of $(a^3/2 e S) P j_0 / v_c = 10^{-4}$ and $v_c \tau_{sd} / \lambda = 0.1$, which are equivalent to the case that, for $v_c \simeq 3 \,[\mathrm{m/s}]$~\cite{yamaguchi2004} and $a \sim 1.5 \, [\AA]$, $j_0  /S \sim 3 \times 10^{8} \, [\mathrm{A/m^2}]$ for $P = 0.1$ and $\tau_{sd} = 6.7 \times 10^{-10} \,[\mathrm{s}]$ assuming $\lambda = 20 \,[\mathrm{nm}]$.
Hence, when observing the DW position as changing the AC frequency, we estimate the exchange interaction strength from the particular frequency in which the oscillation amplitude takes a maximum value.
Note that the current density is four order smaller than common one~\cite{yamaguchi2004}.
\par
In conclusion, we have developed a theory of the interfacial spin-transfer torque and $\beta$-term torque, by consider the bilayer structure of the normal metal and the ferromagnet with the spatially-varying magnetic texture, applying alternating current parallel to the interface.
We find that both torques are enhanced as the alternating current frequency $\omega$ approaches to $1 / \tau_{sd} = 2 \Delta / \hbar$.
We also find that the $\beta$-term torque we obtain here includes a novel contribution which is proportional to the time derivative of the current and exists even in the absence of spin relaxation processes.
Evaluating the domain wall motion due to the spin torques, we directly estimate the interfacial exchange interaction strength.
We have revealed an aspect of the spin-transfer torque with finite frequency, which are enhanced by the resonance of electronic states.
By using this enhancement, less current density is needed to magnetization dynamics, which may lead to low-energy consuming magnetic devices.
\par
We thank S.~Maekawa, G.~Tatara, and J.~Shibata for fruitful discussion.
We also thank Y.~Nozaki for giving stimulating information.

\bibliography{reference}
\end{document}